\documentstyle[aps,prl,psfig,epsf,floats]{revtex}
\begin{document}
\draft
\twocolumn[\hsize\textwidth\columnwidth\hsize\csname@twocolumnfalse%
\endcsname \title{Finite-Temperature Phase Diagram of Hard-Core Bosons in Two
  Dimensions} \author{Guido Schmid$^{(1)}$, Synge Todo$^{(1,2)}$, Matthias
  Troyer$^{(1)}$, and Ansgar Dorneich$^{(3)}$} \address{$^{(1)}$Theoretische
  Physik, Eidgen\"ossische Technische Hochschule Z\"urich, CH-8093 Z\"urich,
  Switzerland \\ $^{(2)}$Institute for Solid State Physics, University of
  Tokyo,
  Kashiwa 277-8581, Japan \\
  $^{(3)}$Institut f\"ur Theoretische Physik, Universit\"at W\"urzburg, 97074
  W\"urzburg, Germany} \date{\today} \maketitle
\begin{abstract}
  We determine the finite-temperature phase diagram of the square lattice
  hard-core boson Hubbard model with nearest neighbor repulsion using quantum
  Monte Carlo simulations.  This model is equivalent to an anisotropic
  spin-1/2 $XXZ$ model in a magnetic field.  We present the rich phase diagram
  with a first order transition between solid and superfluid phase, instead of
  a previously conjectured supersolid and a tricritical end point to phase
  separation. Unusual reentrant behavior with ordering upon {\it increasing}
  the temperature is found, similar to the Pomeranchuk effect in $^3$He.
\end{abstract}
%PACS numbers: 74.25Dw, 75.10Jm, 67.40Kh, 05.30Jp}
\vskip-0.4cm
]
\renewcommand{\floatpagefraction}{0.99} A nearly universal feature of strongly
correlated systems is a phase transition between a correlation-induced
insulating phase with localized charge carriers, to an itinerant phase.  High
temperature superconductors \cite{hightc}, manganites \cite{manganites} and
the controversial two dimensional (2D) ``metal-insulator-transition'' 
\cite{mit} are just a few
examples of this phenomenon. The 2D hard-core boson Hubbard
model provides the simplest example of such a transition from a
correlation-induced charge density wave insulator near half filling to a
superfluid (SF). It is a prototypical model for preformed Cooper pairs
\cite{Siller}, of spin flops in anisotropic quantum magnets \cite{AvO,Kohno},
of SF Helium films\cite{helium} and of supersolids \cite{Batrouni1,Batrouni2}.

In simulations on this model, which does not suffer from the negative sign
problem of fermionic simulations, we can investigate some of the pertinent
questions about such phase transitions: what is the order of the quantum phase
transitions in the ground state and the finite temperature phase transitions?
Are there special points with dynamically enhanced symmetry
\cite{FisherNelson}?  Can there be coexistence of two types of order (such as
a supersolid -- coexisting solid and superfluid order)?  Answers to these
questions provide insight also for the other problems alluded to above.

The Hamiltonian of the hard-core boson Hubbard model we study is
\begin{eqnarray}
H=&-&t\sum_{\langle
    {\bf i},{\bf j}\rangle}(a_{{\bf i}}^{\dagger}a_{{\bf j}}+
a_{{\bf j}}^{\dagger}a_{{\bf i}})
+V\sum_{\langle {\bf i},{\bf j} \rangle} n_{{\bf i}} n_{{\bf j}}
-\mu \sum_{\bf i} n_{\bf i},
\label{HCBham}
\end{eqnarray}
where $a_{\bf i}^{\dagger}$ $(a_{\bf i})$ is the creation (annihilation)
operator for hard-core bosons, $n_{\bf i}=a_{\bf i}^{\dagger} a_{\bf i}$ the
number operator, $V$ the nearest neighbor Coulomb repulsion and $\mu$ the
chemical potential. This model is equivalent to an anisotropic spin-1/2 $XXZ$
model with $J_z=V$ and $|J_{xy}|=2t$ in a magnetic field $h=2V-\mu$. The zero
field (and zero magnetization $m_z=0$) case of the spin model corresponds to
the half filled bosonic model (density $\rho=\langle m_z \rangle=1/2$) at
$\mu=2V$.  Throughout this Letter we will use the bosonic language, and refer
to the corresponding quantities in the spin model where appropriate.  Due to
the absence of efficient Monte Carlo algorithms for classical magnets in a
magnetic field there are still many open questions even in the classical
version of this model, which was only studied by a local update method
\cite{Landau}.

\begin{figure}
  \psfig{file=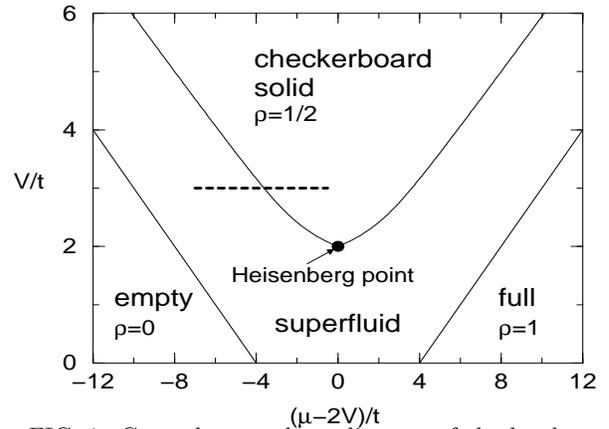,height=2.29in,width=3.1in} \vskip-2mm
\caption{Ground-state phase diagram of the hard-core boson Hubbard model.
  The dashed line indicates the cut along which we calculated the
  finite-temperature phase diagram shown in Fig.~\ref{fig2}.}
\label{fig:groundstate}
\end{figure}

In Fig.~\ref{fig:groundstate} we show the
ground-state phase diagram \cite{Kohno,Batrouni2,Hebert}.  For 
dominating chemical potential $\mu$ the
system is in a band insulating state ($\rho=0$ and $\rho=1$ respectively),
while it shows staggered checkerboard charge order ($\rho=1/2$) for dominating
repulsion $V$.  These solid phases are separated from each other by a SF.
Earlier indications for a region of supersolid phase between checkerboard
solid and SF phase turned out to be due to phase separation at this transition
which is of first order at $T=0$ \cite{Kohno,Batrouni2,Hebert}.

All of these phases extend to finite temperatures.  On the strong repulsion
side the hard-core boson Hubbard model is equivalent to an antiferromagnetic
Ising model at $t=0$ and the insulating behavior extends up to a
finite-temperature phase transition of the Ising universality class ($T_c
\approx 0.567V$ at half filling).  In case of zero repulsion the system is
equivalent to an XY model in a magnetic field and the SF extends up to a
transition of Kosterlitz Thouless (KT) type at finite temperature.  The point
$V=2t$ and $\mu=2V$ is also of special interest, it corresponds to the
Heisenberg point in the spin model and has an enlarged $O(3)$ symmetry, with
long range order only at zero temperature in 2D \cite{Mermin}.

We focus our attention on the finite-temperature phase diagram in the vicinity
of the first order quantum phase transition separating the SF and solid ground
states away from the Heisenberg point. It is instructive to first consider the
three dimensional (3D) version of this model, with a similar ground-state
phase diagram. There, Fisher and Nelson \cite{FisherNelson} determined that
the first order line terminates at a finite-temperature bicritical point at
which the transition temperatures of the solid and SF phases meet.  This
bicritical point has a dynamically enhanced $O(3)$ symmetry -- the symmetry of
the Hamiltonian is only $O(2)$.

This feature alone dictates that despite the similar ground-state phase
diagrams, the finite-temperature phase diagrams have to be very different in
2D, as an $O(3)$ critical point in 2D can exist only at zero temperature
\cite{Mermin}. Even with the absence of a supersolid in the ground state
several alternative scenarios are still plausible: (i) the bicritical point
could be pulled down to zero temperature, and the transition temperatures
$T_{KT}$ and $T_c$ approach zero temperature without a direct phase transition
between SF and solid at finite temperatures, (ii) a supersolid phase could
exist at finite temperatures, although it is absent in the ground state, or
(iii) the phase transition lines meet at finite temperature, but without
dynamically enhanced symmetries.

\begin{figure}
  \psfig{file=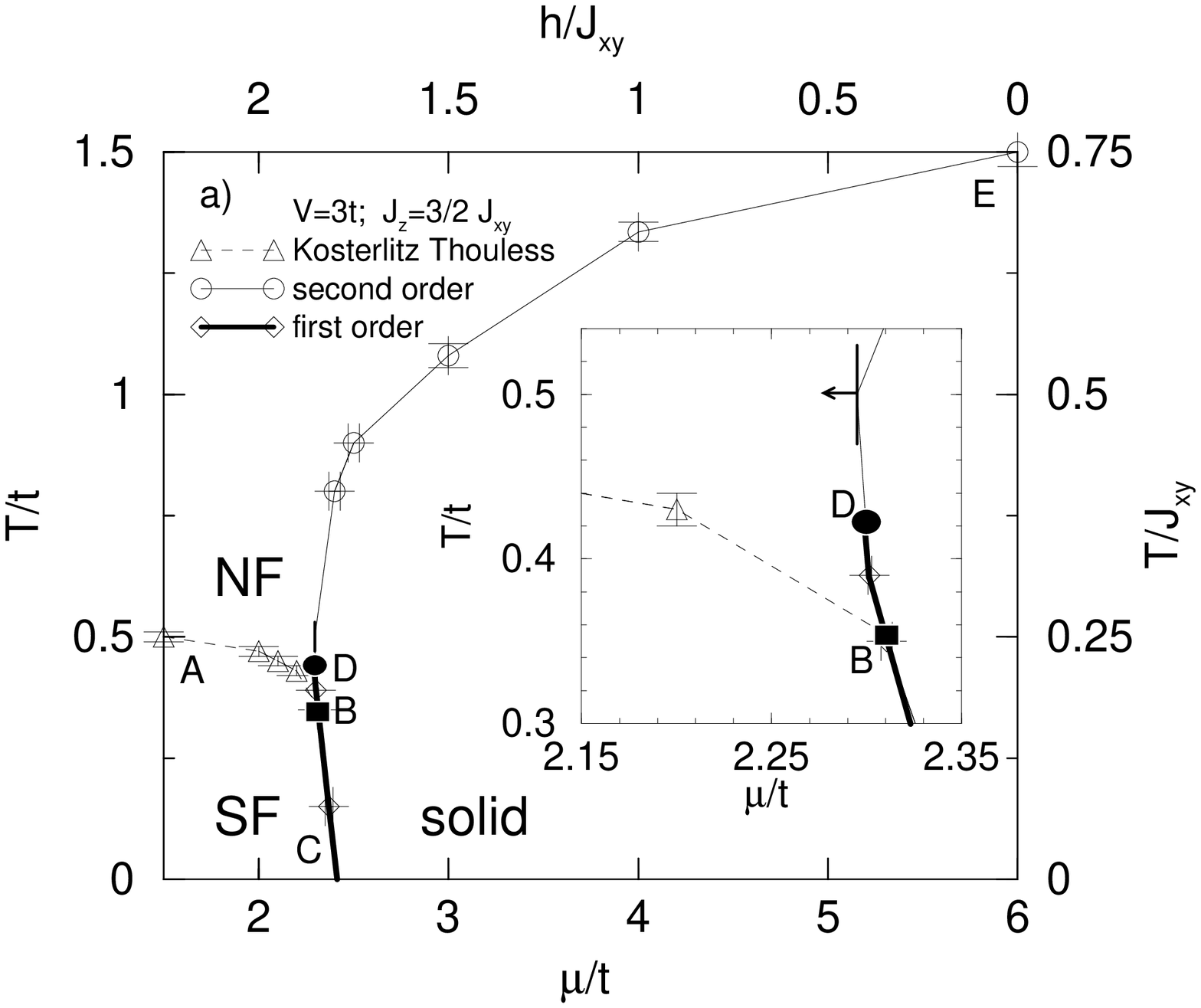,height=2.66in,width=3.1in} \vskip-2mm
  \psfig{file=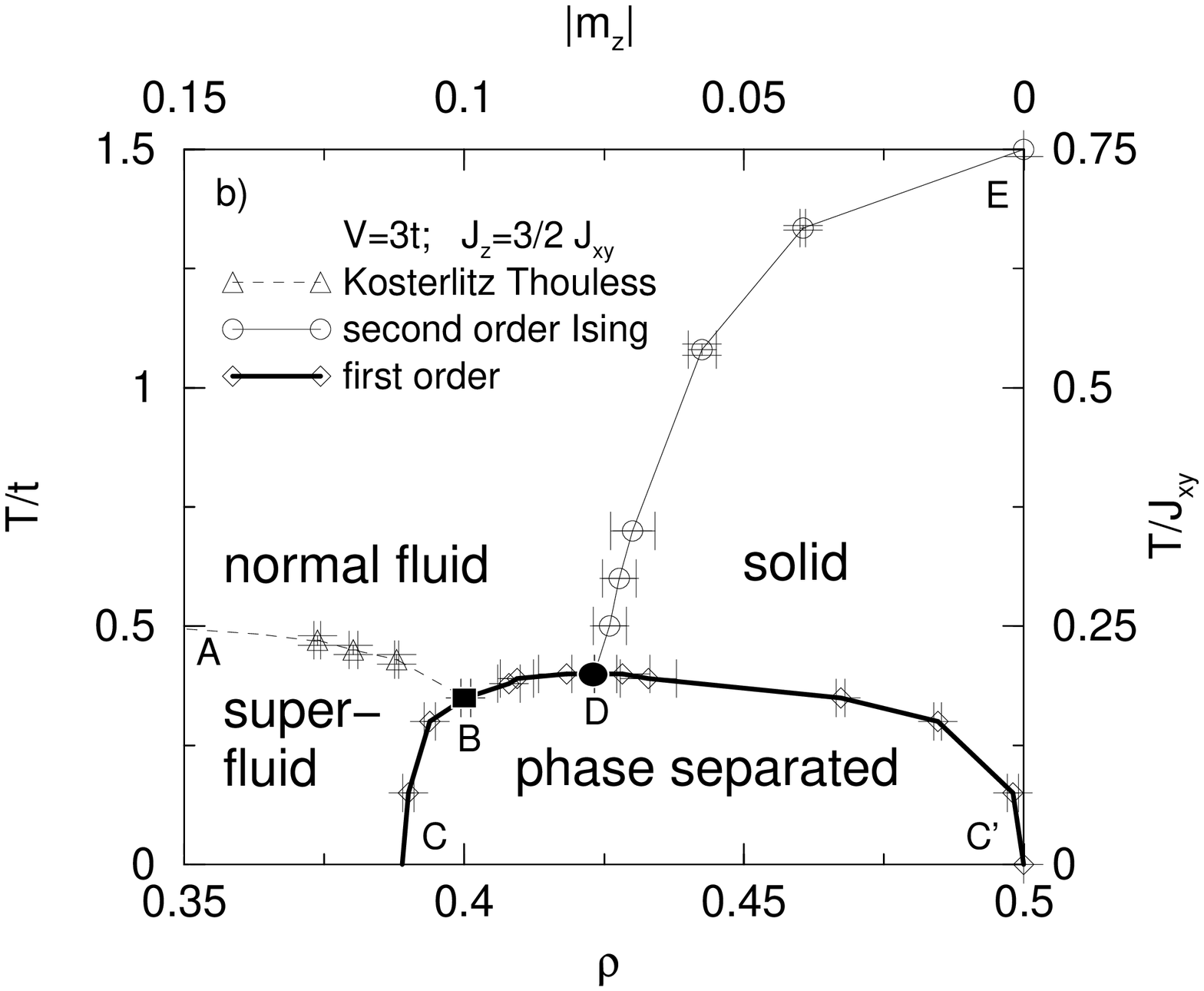,height=2.66in,width=3.1in}
\caption{Phase diagrams at $V=3t$ as function of a) $\mu$ and $T$ and 
  b) $\rho$ and $T$. The arrow in the inset of the
  upper panel denotes an upper bound for the critical $\mu$ at $T=0.5t$. The
  error bars of the triangles are strict bounds for the KT transition. The
  right and upper axis show the parameters in the spin language. The solid
  circle and square denote qualitatively where we expect the critical points
  to be. Lines are guides to the eye. }
\label{fig2}
\end{figure}

In order to decide between these possibilities and determine the
finite-temperature phase diagram we use quantum Monte Carlo simulations,
employing a variant \cite{Sandvik,Dorneich} of the worm algorithm \cite{worm}
in the stochastic series expansion (SSE) representation \cite{Sandvik}. The
simulations were performed in the grand canonical ensemble using lattices up
to size $96\times96$ and do not suffer from any systematic error apart from
finite size effects. In contrast to the loop algorithm \cite{loop} the worm
algorithm, while not as efficient at half filling (a spin model in zero
magnetic field), remains effective away from half filling (in a finite field),
where the loop algorithm slows down exponentially.

From now on we will restrict our discussion to $V=3t$, which corresponds to
the dashed line in Fig.~\ref{fig:groundstate}.  We present the
finite-temperature phase diagram for this representative value in
Fig.~\ref{fig2}. Due to particle-hole symmetry it is sufficient to show
densities up to $1/2$ and chemical potentials up to $\mu=2V=6t$. We find that
scenario (iii) is realized: the first order phase transition extends to a
tricritical point (D) at around $T_3\approx 0.4t$, where it meets with the
second order melting transition of the solid. The KT transition temperature
$T_{KT}$ of the SF meets the first order line at a lower temperature (B) -- in
contrast to the 3D case, where all the three lines meet at the same bicritical
point (B$=$D) with enhanced $O(3)$ symmetry. We want to emphasize one unusual
feature before discussing in detail how the phase diagram was determined:
there is reentrant behavior in the vicinity of the tricritical point (D) and
the system melts upon {\it decreasing} the temperature at constant chemical
potential.

\begin{figure}
  \psfig{file=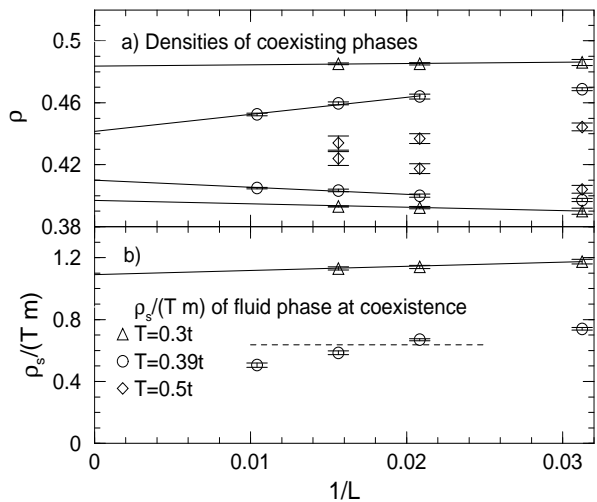,height=2.66in,width=3.1in} \vskip-00mm
\caption{a) Finite size scaling of the coexistence densities at 
  different temperatures, demonstrating the first order nature of the phase
  transition at $T\le0.39t$. b) Finite size scaling of the SF density $\rho_s$
  at different temperatures, showing that the first order transition is
  between a solid and SF at low temperatures, and between a solid and normal
  fluid at higher temperatures.  The dashed line shows the value of $\rho_s/(T
  m)$ at a KT transition.  All simulations are done at $V=3t$.}
\label{fig4}
\end{figure}

The first order phase transition is characterized by a finite jump
$\Delta\rho$ in the density. As discussed in Ref. \cite{Hebert}, where we
present the ground-state phase diagram, two peaks can be observed in the
density distribution at the critical chemical potential. Since a two-peak
structure can also be seen at a second order transition in a finite system, a
careful finite size scaling analysis is needed. Figure~\ref{fig4}a shows the
finite size scaling of the densities of the two coexisting phases. For
$T\le0.39t$ the jump $\Delta\rho$ remains finite under extrapolation, thus
confirming the first order nature up to this temperature (line from C to D in
Fig.~\ref{fig2}), while it scales to zero for $T\ge0.45t$, an indication for a
second order transition (line from D to E).  Additionally we have checked the
finite size scaling of the fourth order cumulant ratio $C_4=1-\langle
m_s^4\rangle/3\langle m_s^2\rangle^2$, of the staggered charge order parameter
$m_s$ in the solid phase. Consistency with the expected size independent
universal value of the Ising universality class $C_4^c=0.6106900(1)$
\cite{kamieniarz} at the transition confirms the second order nature. We have
thus determined that the first order transition extends up to a critical point
(D) at a temperature larger than $0.39t$ and smaller than $0.45t$. In the next
paragraph we show that the critical point where the KT line meets the first
order line (point B) is below $0.39t$. Therefore the end point D of the first
order line is a {\it tri}critical point.

The KT transition temperatures $T_{KT}$ (line from A to B in Fig.~\ref{fig2})
can be determined using the SF number density $\rho_s=(mT/ 2)\langle
W_x^2+W_y^2 \rangle$ \cite{harada}.  Here $W_x$ and $W_y$ are the winding
numbers of the world lines in $x$ and $y$ direction respectively, and
$m=1/(2t)$ is the mass of a boson. For system size $L = \infty$, $\rho_s$ has
a universal value at $T_{KT}$, $\rho_s (T_{KT}) = (2/\pi)mT_{KT}$, while it is
always larger than this universal value for $T<T_{KT}$.  At $T=T_{KT}$,
$\rho_s$ jumps from the universal value to zero and we have $\rho_s(T)=0$ for
all $T>T_{KT}$.  However, for finite $L$, $\rho_s$ is nonzero at all $T$ and
the finite size corrections of $\rho_s$ at $T=T_{KT}$ are given by
\cite{harada} $\rho_s \pi=2mT_{KT}(1+\{2\log[L/L_0(T_{KT})]\}^{-1})$.
Although these finite size corrections are known to be notoriously difficult,
we can obtain strict upper and lower bounds for $T_{KT}$ by plotting $1/(\pi
\rho_s /mT -2) - \log(L)$ as function of $\log L$. As $L$ is increased, this
quantity converges to the constant $(-\log L_0 )$ at $T_{KT}$ and diverges to
$\pm \infty$ for $T>T_{KT}$ and $T<T_{KT}$, respectively \cite{troyer}. The
error bars we obtain using this method are reliable and small enough for our 
purposes.

To investigate how the KT transition line meets the phase separation line we
perform canonical measurements of the SF density $\rho_s$ at the density
$\rho_f$ of the fluid on this coexistence line. In Fig.~\ref{fig4}b we show
the finite size scaling of $\rho_s(\rho_f)$. At low temperatures $T=0.3t$
$\rho_s$ of the SF at the coexistence line is well above the universal value
for a KT transition. Therefore the transition between SF and solid at
temperatures $T\le 0.3t$ does not include a KT transition. The system remains
SF right up to a direct first order phase transition from the SF to the solid.
At $T=0.39t$ on the other hand, $\rho_s$ as a function of the system size $L$
drops below the universal value and should scale to zero according to
\cite{jump}. There is thus a first order transition from solid to normal fluid
at $T=0.39t$ and the KT line has to meet the first order line at a finite
temperature between $0.3t$ and $0.39t$.

\begin{figure}
  \psfig{file=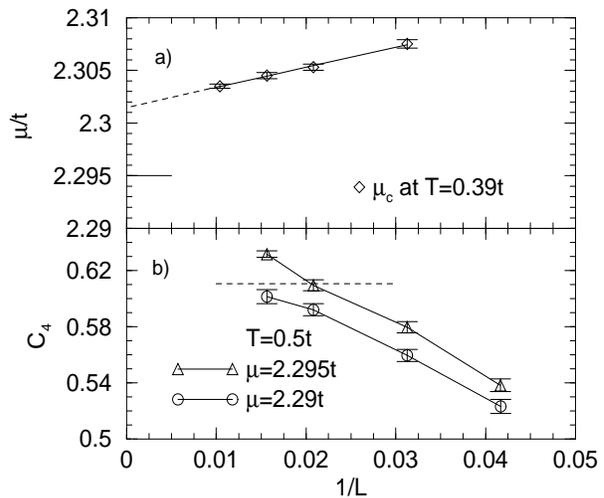,height=2.66in,width=3.1in} \vskip-00mm
\caption{a) Finite size scaling of the critical chemical potential 
  $\mu_c$ at $T=0.39t$.  The horizontal line is an upper bound for the
  critical chemical potential at $T=0.5t$. b) Finite size scaling of the
  fourth order cumulant $C_4$ at $T=0.5t$ for different chemical potentials.
  The horizontal line is the universal value of $C_4$ at the Ising critical
  point. All simulations are done at $V=3t$.}
\label{fig5}
\end{figure}

As can be seen in Fig.~\ref{fig2}, the checkerboard solid is melting in a
first order transition to a SF upon decreasing the temperature around
$\mu\simeq2.35t$.  Even more special is the melting to a normal fluid in the
vicinity of the tricritical point (D) both upon increasing or decreasing the
temperature.  To proof this reentrant behavior, we first determine the
critical chemical potential $\mu_c(T=0.39t)$ of the first order transition at
$T=0.39t$. We then show that $\mu_c(T=0.5t)<\mu_c(T=0.39t)$ On a finite system
we define $\mu_c$ as the chemical potential where there is an equal
probability to be in the solid and fluid phase (equal peak areas in the
density probabilities). Performing a finite size extrapolation
(Fig.~\ref{fig5}a) we obtain $\mu_c(0.39t)=2.3015(10)t$.  Next we calculate
the upper bound for the critical chemical potential $\mu_c(T)$ at $T=0.5t$.
Observing that the fourth order cumulant ratio $C_4$ scales above the
universal critical value $C_4^c$ at $\mu=2.295t$ (Fig.~\ref{fig5}b), we thus
have $\mu_c(T=0.5t)<2.295t$. As $\mu_c(T=0.5t)<\mu_c(T=0.39t)$ we have
pinpointed reentrant behavior where the solid melts to a normal fluid both
when increasing and decreasing the temperature.

To summarize, we have investigated the finite-temperature phase diagram of the
hard-core boson Hubbard model with nearest neighbor repulsion using quantum
Monte Carlo simulations.  The first order phase transition in the ground state
between SF and solid phases extends to finite temperatures and turns into a
second order melting line of the solid phase at a finite-temperature
tricritical point. As the KT line meets the first order line at a lower
temperature, we have a qualitatively different phase diagram than in 3D
\cite{FisherNelson}, where there is a bicritical point with dynamically
enhanced $O(3)$ symmetry.

We expect this phase diagram, obtained for a representative value of $V=3t$,
to be valid also for other couplings $V>2t$. On approaching the Heisenberg
point $V=2t$ the temperature scale of all the transitions drops to zero.  In
the other limit $t/V\rightarrow0$, we approach the antiferromagnetic Ising
model.  The KT transition temperatures $T_{KT}$ and the tricritical point
$T_3$ should scale with $t$, since in the Ising model the phase transition is
first order at zero temperature, but second order at any finite temperature
\cite{Todo}, consistent with $T_3/V \sim t/V \rightarrow0$.

The melting line of the solid shows several unusual features. As a function of
chemical potential there are three different phase transitions: at low
temperatures the melting transition is first order into a SF. At intermediate
temperatures the melting transition is still first order, but to a normal
fluid, while it is a {\it second order} transition to a normal fluid at even
higher temperatures.

There is reentrant behavior in the vicinity of the tricritical point: as a
function of increasing temperature there are three phase transitions: first
the SF turns into a normal fluid.  Upon {\it increasing} the temperature
further this normal fluid {\it solidifies}, and finally melts again. The
reason for this reentrant behavior is conjectured to be a large entropy of the
weakly interacting vacancies (or interstitials at $\rho>1/2$) in the solid,
larger than the entropy of the normal fluid.  Such an ordering transition upon
{\it increasing} temperature is rare but has been observed
before, e.g. in the bcc Ising antiferromagnet \cite{reentrant}.

Finally, we want to mention the striking qualitative similarity of the
presented phase diagram to those of fermionic $^3$He and bosonic $^4$He in two
dimensions.  In $^3$He there is also a reentrant melting line (the Pomeranchuk
effect \cite{He3}) and a SF phase at low temperatures \cite{He3Review}. In
$^3$He however the large entropy of the solid is due to the nuclear spin
degrees of freedom and the effect even more pronounced. In $^4$He films, the
solid phase melts to a SF in a small range for the pressure upon {\it
  decreasing} the temperature \cite{gordilloceperley}, like in our phase
diagram where there is a direct transition form SF to solid for constant
values of $\mu$ around $\mu\simeq 2.35t$. Note that in our phase diagram we
have pinned down the even more subtle effect of a reentrant melting line
between solid and normal fluid.

We wish to thank G.G. Batrouni and D.M. Ceperley for stimulating discussions.
The simulations were performed on the Asgard Beowulf cluster at ETH Z\"urich,
employing the Alea library for Monte Carlo simulations \cite{alea}.  G.S.,
S.T. and M.T.  acknowledge support of the Swiss National Science Foundation.

\end{document}